# Square Kilometer Array Telescope – Precision Reference Frequency Synchronisation via 1f-2f Dissemination

B. Wang[1,2], X. Zhu[1,3], C. Gao[1,2], Y. Bai[1,3], J.W. Dong[1,2], and L. J. Wang[1,2,3,4,*]

[1]Joint Institute for Measurement Science, Tsinghua University, Beijing 100084, China

[2]The state key lab of precision Measurement Technology and Instrument, Department of Precision Instruments, Tsinghua University, Beijing 100084, China

[3]Department of Physics, Tsinghua University, Beijing 100084, China

[4]National Institute of Metrology, Beijing 100013, China

* Corresponding authors: lwan@tsinghua.edu.cn

**The Square Kilometer Array (SKA) is an international effort to build the world's largest radio telescope, with one square kilometer collecting area. Besides its ambitious scientific objectives, such as probing the cosmic dawn and cradle of life, SKA also demands several revolutionary technological breakthroughs, with ultra-high precision synchronisation of the frequency references for thousands of antennas being one of them. In this report, aimed at applications to SKA, we demonstrate a frequency reference synchronization and dissemination scheme with the phase noise compensation function placed at the client site. Hence, one central hub can be linked to a large number of client sites, forming a star-shaped topology. As a performance test, the 100 MHz reference signal from a Hydrogen maser clock is disseminated and recovered at two remote sites. Phase noise characteristics of the recovered reference frequency signal coincides with that of the hydrogen-maser source and satisfies SKA requirement.**

Benefiting from innovations in modern atomic clock, time and frequency have been the most accurate and stable physical quantities that can be measured and controlled[1-4]. With significant progress in precision time and frequency synchronisation and dissemination[5-12], initially aimed for metrology[13], more and more areas can share these powerful tools and covert the measurement of target parameters to that of time and frequency. For example, in the applications of radio telescope array[14-16], the spatial resolution is converted to the phase resolution of received signals. In these applications, high-precision, short term phase (frequency) synchronisation (1s to 100 s) between different telescope dishes is required. This requirement normally cannot be satisfied by traditional synchronisation method via satellite links, which can only realize long term (>1000 s) frequency synchronisation[17,18]. Furthermore, satellite based time and frequency synchronisation cannot meet the radio silence requirement of radio astronomy observation. In this regard, recently developed fibre-based frequency synchronisation method is more suitable and can realize ultra-high precision phase (frequency) synchronisation at the integration time from 1 to $10^6$ s[19]. To increase its availability and accessibility, several fibre-based multiple-access frequency dissemination schemes have been proposed and demonstrated[20-25]. However, there are still many unsolved problems.

Specific to applications of SKA[14], there are three main requirements on the fibre-based frequency reference synchronisation system. First, the synchronisation system should distribute the frequency reference signal to hundreds of receivers in a star topology during the first phase of SKA (SKA1). This requirement will be extended to thousands of receivers in SKA2. Secondly, according to the coherence requirement of reference frequency signals, the frequency dissemination stability should be better than $1\times10^{-12}$/s, $1\times10^{-13}$/10s and $1\times10^{-14}$/100s, respectively. And finally, according to the jitter requirement of recovered sampling clock at each receptor, the single-side-band (SSB) phase noise of reference frequency signal should not degrade through dissemination. These requirements cannot be well satisfied by current fibre-based frequency dissemination schemes with active compensation. For all current schemes, the phase noise detection and compensation functions are normally located at the transmitting site. In the case of SKA1 where hundreds of telescope dishes are planned, corresponding number of compensation modules need to be placed in the same central station. This would generate extraordinary space requirement, and cause unnecessary complexities and difficulties for future expansion. Furthermore, the phase noise compensation bandwidth is related to the propagation delay of light in the fibre, which will cause the so called delay-unsuppressed fibre noise.

In this report, we propose and demonstrate a new frequency synchronization and dissemination method with the phase noise compensation function placed at the client (dish) site. One central transmitting module can thus be linked with multiple client sites, and the expansion of future receiving sites will not disrupt the structure of the central transmitting station. As a performance test, using two 50 km fibre spools, we recover the 100 MHz disseminated reference frequency signals received at two separate remote sites. Relative frequency stabilities between the two recovered frequency signals of $3.7 \times 10^{-14}$/s and $3.0 \times 10^{-17}/10^5$s are obtained and far exceeds that of the SKA requirement. Furthermore, the SSB phase noise of the recovered 100 MHz signal is measured and coincides with that of the hydrogen maser (H-maser) clock's source signal at low frequency (<60 Hz frequency offset), and is better at higher frequency offset. The proposed scheme can well-satisfy all frequency synchronisation and dissemination requirements of SKA.

**Results**

**Client-side, 1f-2f actively compensated frequency dissemination method.** Figure 1 shows the schematic diagram of the client-side actively compensated frequency dissemination system. One transmitting site (TX) simulating the central station is connected to two receiving sites (RX). As a performance test and for the convenience of phase difference measurement, the TX is linked with RX-I and RX-II via two 50 km fibre spools, respectively. In our case, different with current active phase compensation schemes, the function of TX is very simple - modulating and broadcasting. To increase the signal-to-noise ratio for compensation, the 100 MHz reference frequency ($V_{ref}$ from a H-maser clock) is boosted to 2 GHz via a phase-locked dielectric resonant oscillator (PDRO), which can be expressed as $V_0 = \cos(\omega_0 t + \phi_0)$ (without considering its amplitude). $V_0$ is used to modulate the amplitude of a 1547.72 nm diode laser. Taking the dissemination channel TX to RX-I as an example, after passing a "2 to 1" fibre coupler and an optical circulator, the modulated laser light is coupled into the 50 km fibre link. The structure of RX is more complex. A 1 GHz PDRO, which can be expressed as $V_1 = \cos(\omega_1 t + \phi_1)$, is phase locked to a 100MHz oven-controlled crystal oscillator (OCXO). The phase of OCXO can be controlled by an external voltage. $V_1$ is used to modulate the amplitude of a 1548.53 nm diode laser. With the help of two optical circulators at RX-I and TX sites, the modulated 1548.53 nm laser light propagates through the same 50 km fibre spool via the route RX-I to TX and back. After the round trip propagation, the 1548.53 nm laser light can be separated from the received one-way 1547.72 nm laser light by a wavelength division multiplexer (WDM). The modulated 1547.72 nm laser is then detected by a fast photo-diode FPD1. The directly received 2 GHz frequency signal from TX can be expressed as $V_2 = \cos(\omega_0 t + \phi_0 + \phi_p)$, where $\phi_p$ is the phase fluctuation induced during the 50 km fibre dissemination. The modulated 1548.53 nm laser carrier is detected by FPD2 and the recovered 1 GHz frequency signal can be expressed as $V_3 = \cos(\omega_1 t + \phi_1 + \phi'_p)$, where $\phi'_p$ is its phase fluctuation induced by the 100 km fibre dissemination (round trip). We then mix-down the signals $V_2$ and $V_3$ to obtain $V_4 = \cos[(\omega_0 - \omega_1)t + \phi_0 + \phi_p - \phi_1 - \phi'_p]$. Furthermore, mixing down the signals $V_1$ and $V_4$ gives an error signal $V_5 = \cos[(\omega_0 - 2\omega_1)t + \phi_0 + \phi_p - 2\phi_1 - \phi'_p]$. Finally, $V_5$ is used to close the phase-locked loop (PLL) to control the phase of OCXO. In this case, the relationship of $\omega_0 = 2\omega_1$ is satisfied and under a fiber dispersion condition (discussed later), the one-way phase fluctuation of the 2 GHz frequency signal is the same as the round-trip accumulated phase fluctuation of the 1 GHz frequency signal, $\phi_p = \phi'_p$. Consequently, the error signal can be expressed as $V_5 = \cos(\phi_0 - 2\phi_1)$. When the PLL is closed, the phase of OCXO at RX-I will be locked to the phase of $V_{ref}$ at TX, with $\phi_0 = 2\phi_1$, thus obtaining stable-frequency dissemination from TX to RX-I. As mentioned, the light propagation delay in the fibre link will limit the feedback control time and cause a delay-unsuppressed fibre noise. To ensure the SSB phase noise of client site's 100 MHz OCXO is not worsen by the unsuppressed fibre noise, we use a self-designed narrow bandwidth PLL. Its bandwidth can be tuned and set based on the phase noise characteristics of the free-running OCXO and H-maser clock. In our case, the PLL's bandwidth is set as 10 Hz. The same structure is applied to the RX-II and all other future sites. The method is referred to as the "1f-2f" method due to the factor-2 relationship between the TX and RX frequencies.

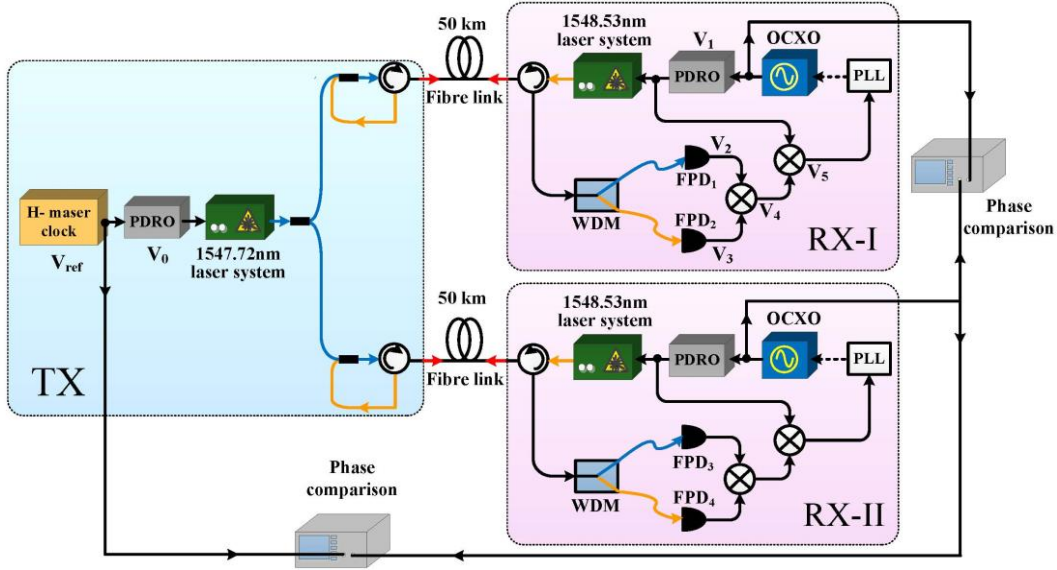

Figure-1 (Wang)

**Figure 1 | Schematic diagram of the client side, 1f-2f actively compensated frequency dissemination system.** PDRO: phase-locked dielectric resonant oscillator, OCXO: oven-controlled crystal oscillator. WDM: wavelength division multiplexer. FPD: fast photo-diode.

**Frequency dissemination stability measurement.** To test the dissemination stability of the proposed client-site active compensation method, we perform a series of measurement. The results are shown in Figure 2. Curve (a) is the measured relative stability of 100 MHz frequency signals between the H-maser at TX and OCXO at RX-I when the PLL is open. The relative frequency stabilities of $1.6 \times 10^{-11}$/s and $1.2 \times 10^{-9}/10^5$s reflect the frequency stability of the OCXO we used. Curve (b) shows the relative frequency stability between $V_0$ and $V_2$. The relative frequency stabilities of $2.0 \times 10^{-12}$/s and $3.4 \times 10^{-14}/10^5$s represent the 50 km fiber dissemination stability without compensation. Curve (c) is the measured relative stability of 100 MHz frequency signals between the H-maser clock at TX and OCXO at RX-I when the PLL is closed with 3Hz effective bandwidth. The relative frequency stabilities of $3.1 \times 10^{-14}$/s and $2.7 \times 10^{-17}/10^5$s are obtained, which means the OCXO at RX-I has been phase locked to the H-maser at TX. We also measure the relative frequency stability between the two OCXOs at RX-I and RX-II when the phase-lock loops at both dissemination links are closed, as shown in curve (d). The relative frequency stabilities of $3.7 \times 10^{-14}$/s and $3.0 \times 10^{-17}/10^5$s have been realized, which means the OCXOs at two RXs are phase coherent to each other within the compensation bandwidth of the dissemination system.

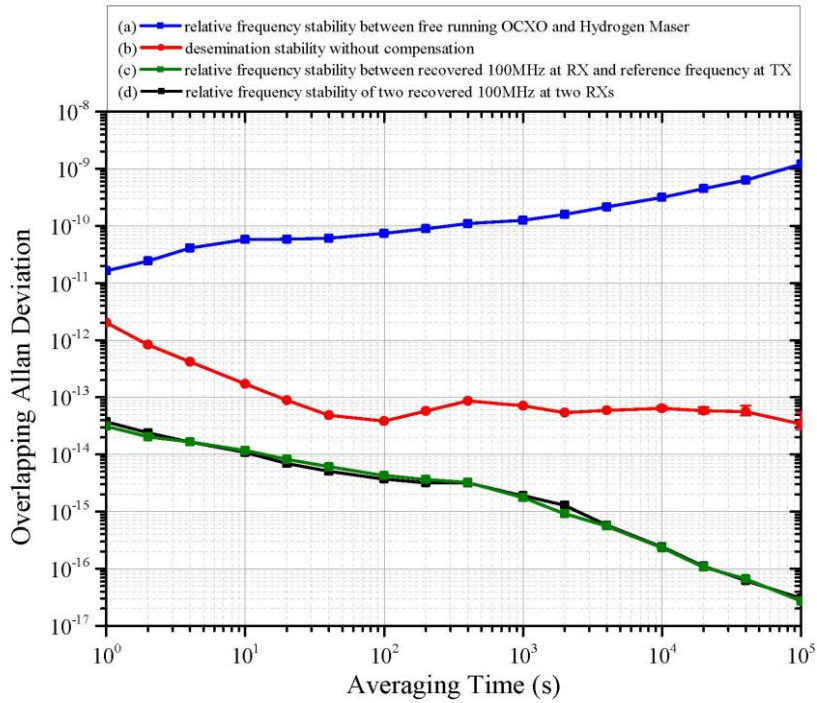

Figure-2 (Wang)

**Figure 2 | Relative frequency stability measurement results.** (a) Relative frequency stability between free running OCXO and H-Maser; (b) Measured frequency stability of the dissemination system without compensation; (c) Relative frequency stability between recovered 100MHz at RX and $V_{ref}$ at TX when the PLL is closed; (d) Relative frequency stability of two recovered 100MHz at two RXs when two PLLs are closed. Curves (c) and (d) are measured simultaneously.

**SSB phase noise measurements.** Above measurements have tested the frequency dissemination stability from 1 s to $10^5$ s integration time, in other words, it is the long term stability (longer than 1 s) of the recovered 100 MHz frequency signal. In addition, outstanding short term stability (shorter than 1 s) for the recovered 100 MHz frequency signal is also required by SKA. Figure 3 shows the measured 100 MHz signals' SSB phase noise spectra in different cases. Cure (a) and (b) are measured SSB phase noise spectrums of the H-Maser clock and free-running OCXO we used, respectively. Obviously, the long term phase stability of free-running OCXO is worse than that of H-Maser clock, its high frequency offset (>60 Hz) phase noise characteristics is better than that of the H-Maser clock. The main crossing point of these two spectra occurs at around 10 Hz. Curve (c) shows the SSB phase noise of OCXO when the PLL is closed under a wideband locking mode. In this case, the loop bandwidth is decided by the light propagation delay, and can be calculated via $1/4\tau \sim 500$ Hz, where $\tau$ is the round-trip propagation time of light in the 50 km fiber link. Consequently, there is a big bump around 500 Hz frequency offset on curve (c). Furthermore, the unsuppressed high frequency phase noise (>500 Hz) induced during the fiber dissemination worsen the phase-locked OCXO's phase noise. Curve (d) shows the SSB phase noise of OCXO when the effective bandwidth of PLL is set at 10 Hz. In this case, its SSB phase noise coincides with that of the H-Maser at below 60 Hz frequency offset, and is better after 60 Hz frequency offset. Based on these measurements, we see that the long term stability of the recovered 100 MHz frequency signal is locked to the H-Maser on the TX site, and its short term phase noise characteristics can even be improved by carefully choosing the locking bandwidth.

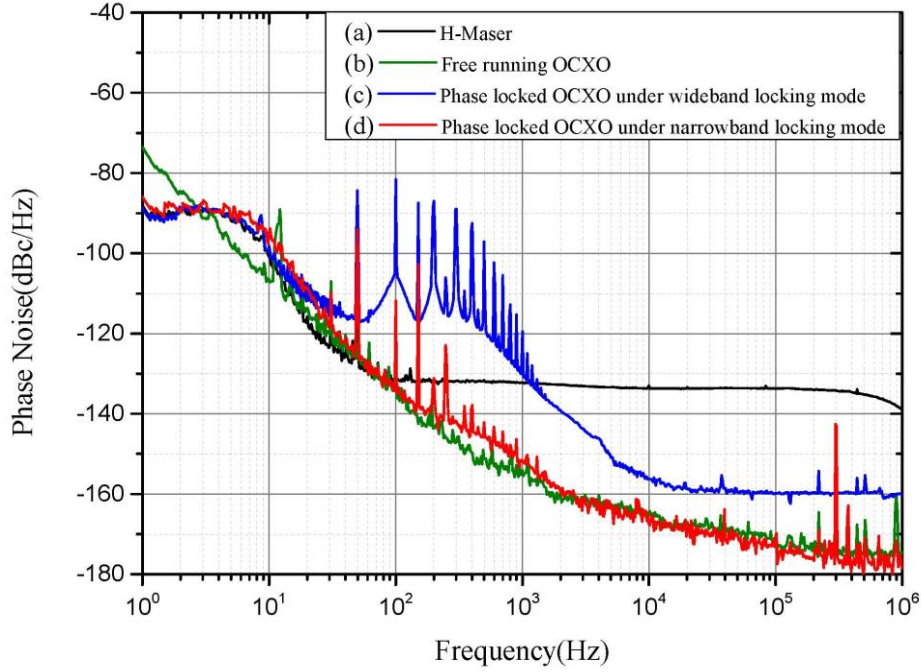

Figure-3 (Wang)

**Figure 3 | Single sideband (SSB) phase noise measurement results of different 100 MHz frequency signals.** (a) The SSB phase noise of H-Maser at TX; (b) SSB phase noise of free-running OCXO used in the experiment; (c) SSB phase noise of the phase-locked OCXO under a wideband locking mode; (d) SSB phase noise of the phase-locked OCXO under narrowband locking mode.

**Discussion**

In the proposed scheme, the two laser carriers' wavelengths are different. Consequently, the fiber dispersion may cause a phase delay difference between the two laser carriers, and this may limit the frequency dissemination stability. Here, we discuss this effect and analyze its effect under a real-world situation. The phase time delay $\tau(\lambda,T)$ of the disseminated frequency signal is

$$\tau(\lambda,T) = \frac{n(\lambda,T) \cdot l}{c}, \qquad (1)$$

where c is the light speed in vacuum, $n(\lambda,T)$ is the fiber refractive index at optical wavelength $\lambda$ and temperature T, L is the fiber length. In practice, temperature fluctuation of the fiber link is the dominant factor which will degrade the frequency dissemination stability. We calculate the second-order partial derivative of phase time delay

$$\begin{aligned}\frac{\partial^2 \tau(\lambda,T)}{\partial \lambda \partial T} &= \frac{1}{c} \cdot \frac{\partial \left[ \frac{\partial n(\lambda,T)}{\partial \lambda} \cdot L + \frac{\partial L}{\partial \lambda} \cdot n(\lambda,T) \right]}{\partial T} \\ &= \frac{\partial \left[ D(\lambda,T) \cdot L \right]}{\partial T} \\ &= L \cdot \frac{\partial D(\lambda,T)}{\partial T} + D(\lambda,T) \cdot \frac{\partial L}{\partial T} \\ &= L \cdot \left[ \kappa + D(\lambda,T) \cdot \alpha \right] \end{aligned} \qquad (2)$$

with the fiber's dispersion

$$D(\lambda,T) = \frac{1}{c} \cdot \frac{dn(\lambda,T)}{d\lambda}, \qquad (3)$$

the chromatic dispersion thermal coefficient

$$\kappa = \frac{dD(\lambda,T)}{dT}, \qquad (4)$$

and the fiber length thermal expansion coefficient being

$$\alpha = \frac{1}{L} \cdot \frac{dL}{dT}. \qquad (5)$$

For the commercial G652 fiber, $\alpha = 5.6 \times 10^{-7} /°C$ [26], $D = 17 \, ps/(nm \cdot km)$ and $\kappa = -1.45 \times 10^{-3} \, ps/(km \cdot nm \cdot °C)$ [27] around 1550 nm. Obviously, the chromatic dispersion thermal coefficient is the dominant item on the phase time delay difference. For a 100 km fiber dissemination, the phase time delay difference will be $\delta \tau = -0.14 \, ps/(nm \cdot °C)$. Supposing a diurnal temperature fluctuation of $30°C$, the phase time delay difference of the frequency signal carried by 1547.72 nm and 1548.53 nm (0.81 nm difference) will be 3.5 ps, which corresponds to the dissemination stability of $8.1 \times 10^{-17}$ at the integration time of half a day. This frequency dissemination stability can satisfy almost all current practical applications. Using the dense wavelength division multiplexing (DWDM) technique, two laser carriers' wavelength can be as near as 0.4 nm, and this will further reduce the chromatic dispersion's impact on the dissemination stability. Consequently, the proposed client-side active compensation scheme can fully satisfy the reference frequency dissemination requirements of SKA. With this scalable scheme, the space requirement and complexity of the frequency dissemination system at the center station can be dramatically reduced.

**Methods**

**Dissemination stability measurement.** As noted, curve (c) and (d) in figure 2 are measured simultaneously. However, only one commercial phase noise measurement device (5125A of Microsemi Corporation) were available. Consequently, we use the so-called "voltmeter" method to measure the dissemination stability with active compensation. During the measurement, two 100 MHz signals are mixed to a DC signal. The DC signal's voltage fluctuation corresponds to the relative phase fluctuation between two 100 MHz signals. It is measured by a precise voltmeter and recorded by a computer. The relative phase time fluctuation can be calculated as $y(t) = \frac{1}{2\pi\nu} \cdot \arcsin(\frac{V(t)}{V_{pp}/2})$. In the formula, $\nu$ is the frequency of comparison signal (here, 100 MHz); $V(t)$ is the measured real-time voltage fluctuation of the DC signal; $V_{pp}$ is the peak to peak voltage of the DC signal when the relative phase change between two comparison signals is large than $2\pi$, it can be measured when the PLL is open. Using the commercial software Stable 32, we can calculate the overlapping Allan deviation via the measured phase time fluctuation. In practice, to enhance the sensitivity of phase fluctuation measurement, $V(t)$ is always set around 0. This can be realized by choosing suitable length of RF cable between OCXO and the phase comparison mixer. Here, the measured result via "voltmeter" method has been verified by the commercial device 5125A. Two results are identical. Compared with 5125A, the "voltmeter" method has its advantage. The frequency measurement range of 5125A is 1-400MHz, while the "voltmeter" method can be used to measure the frequency higher than 400 MHz. The "voltmeter" method's disadvantages are: (1) it cannot be used to measure signals with different frequency; (2) it cannot be used to measure the unstable signals whose relative phase fluctuation is higher than $2\pi$, for example, the curve (a) and curve (b) of figure 2.

**Acknowledgements**

We acknowledge funding supports from the National Key Scientific Instrument and Equipment Development Projects (No.2013YQ09094303).

**Author contributions**

B. Wang and L. J. Wang conceived the experiments and wrote this report. All authors carried out the experiments and contributed to the final manuscript.

**Additional information**

Competing financial interests: The authors declare no competing financial interests.